\documentclass[12pt]{iopart}
\usepackage{iopams}
\usepackage{dsfont}

\newcommand{\half}{\mbox{$\textstyle \frac{1}{2}$}}
\newcommand{\re}{\mbox{$\rm e$}}
\newcommand{\ri}{\mbox{$\rm i$}}
\newcommand{\rd}{\mbox{$\rm d$}}

\begin{document}

\title[Unitarity, ergodicity, and quantum thermodynamics]
{Unitarity, ergodicity, and quantum thermodynamics}

\author[D.~C.~Brody, D.~W.~Hook, \& L.~P.~Hughston]
{Dorje~C.~Brody${}^*$, Daniel~W.~Hook${}^\dagger$, and
Lane~P.~Hughston${}^\ddagger$}

\address{${}^*$Department of Mathematics, Imperial College,
London SW7 2BZ, UK \\  ${}^\dagger$Blackett Laboratory, Imperial
College, London SW7 2BZ, UK \\ ${}^\ddagger$Department of
Mathematics, King's College London, London WC2R 2LS, UK}

\date{\today}

\begin{abstract}
This paper is concerned with the ergodic subspaces of the state
spaces of isolated quantum systems. We prove a new ergodic theorem
for closed quantum systems which shows that the equilibrium state of
the system takes the form of a grand canonical density matrix
involving a complete commuting set of observables including the
Hamiltonian. The result obtained, which is derived for a generic
finite-dimensional quantum system, shows that the equilibrium state
arising from unitary evolution is always expressible in the
canonical form, without the consideration of a system-bath
decomposition.
\end{abstract}

\submitto{\JPA} 

Given the Hamiltonian ${\hat H}$ and the initial state
$|\psi_0\rangle$ of an isolated quantum system, what is the dynamic
average
\begin{eqnarray}
\langle\!\langle {\hat O} \rangle\!\rangle = \lim_{t\to\infty}
\frac{1}{t} \int_0^t \langle\psi_s|{\hat O}|\psi_s\rangle \rd s
\label{eq:1}
\end{eqnarray}
of an observable ${\hat O}$ when the state $|\psi_t\rangle =
\re^{-{\rm i}{\hat H}t} |\psi_0\rangle$ of the system evolves
unitarily? Is there an equilibrium density matrix ${\hat\rho}$, with
a thermodynamic characterisation, such that the average is given by
$\langle\!\langle {\hat O} \rangle\!\rangle={\rm tr}({\hat\rho}
{\hat O})$\,?

In the case of a classical system, if the Hamiltonian evolution is
ergodic, then the theorem of Koopman, von Neumann, and Birkhoff
shows that the dynamic average can be replaced by a statistical
average over a subspace of the phase space determined by the
relevant conservation laws~\cite{khinchin}. If the system consists
of a large number of interacting particles, then the dynamic
average is intractable, whereas the statistical average in many
cases can be calculated.

In the case of quantum systems, while the equilibrium properties of
small subsystems of large systems have been studied
extensively~\cite{kemble,schrodinger,klein,ekstein,bocchieri,
prospert,tasaki,gemmer,goldstein,popescu}, less attention has been
paid to the equilibrium states arising as a consequence of the
unitary evolution of closed systems. The purpose of this paper is to
investigate such systems and to derive rigorous results concerning
(a) the dynamic averages of observables, and (b) the associated
equilibrium states.

We consider an isolated quantum system based on a Hilbert space of
dimension $n+1$, with a generic, nondegenerate Hamiltonian ${\hat
H}$ (the degenerate case will be considered later). We write
$\{E_i\}_{i=0,1, \ldots,n}$ for the energy eigenvalues, and
$\omega_{ij}=E_i-E_j$ for the eigenvalue differences. The normalised
energy eigenstates will be denoted $\{|E_i\rangle\}_{i=0,1,
\ldots,n}$, with the associated projection operators
$\{{\hat\Pi}_i\}_{i=0,1, \ldots, n}$. We write $|\psi_0\rangle$ for
the initial state, and $\{|\psi_t\rangle\}_{0\leq t<\infty}$ for its
unitary evolution under the influence of ${\hat H}$. With these
definitions at hand, the main result can be expressed as follows:

\noindent {\bf Quantum ergodic theorem}. \textit{The dynamic average
of an observable ${\hat O}$ is given by $\langle\!\langle {\hat O}
\rangle\!\rangle = {\rm tr}({\hat\rho}{\hat O})$, where
\begin{eqnarray}
{\hat\rho} = \frac{1}{Z(\beta,\{\mu_i\})}\,\exp\Big(-\beta {\hat H}
- \sum_{i=2}^{n} \mu_i {\hat F}_i\Big), \label{eq:2}
\end{eqnarray}
and $Z(\beta,\{\mu_i\})={\rm tr}\,\exp(-\beta{\hat H}-\sum_{i=2}^n
\mu_i {\hat F}_i)$. Here ${\hat H}$ together with $\{{\hat
F}_i\}_{i=2,\ldots,n}$ constitute a complete set of commuting
observables. The effective inverse temperature $\beta$ and chemical
potentials $\{\mu_i\}_{i=2,\ldots,n}$ are given by the relations
\begin{eqnarray}
\beta=\frac{\partial S}{\partial E}, \quad {\rm and}\quad
\mu_i=\frac{\partial S}{\partial F_i}, \label{eq:3}
\end{eqnarray}
where $E={\rm tr}({\hat\rho}{\hat H})$, and $F_i={\rm tr}(
{\hat\rho}{\hat F}_i)$. The entropy $S=-{\rm tr} ({\hat\rho}
\ln{\hat\rho})$ is given by
\begin{eqnarray}
S=-\sum_{i=0}^n p_i \ln p_i, \label{eq:4}
\end{eqnarray}
with $p_i = |\langle\psi_0|E_i\rangle|^2$}.

The appearance of the grand canonical density matrix (\ref{eq:2}) is
surprising, since this structure normally arises with the
consideration of the equilibrium thermodynamics of a small system
immersed in a thermal bath. Indeed, the canonical form
${\hat\rho}=\exp(-\beta {\hat H})/Z(\beta)$ is known to appear in
the case of a system in a thermal bath for an overwhelming majority
of wave functions of the total system~\cite{goldstein,popescu}.
Equation (\ref{eq:2}) is a stronger result, valid in the case of a
closed system, involving no approximations and no invocation of the
thermodynamic limit.

To determine the equilibrium states of a closed quantum system we
need to identify the subspaces of the quantum state space over which
a generic time evolution will exhibit ergodicity. The idea is that
in general there are $n$ conserved quantities arising in connection
with unitary evolution in a Hilbert space of dimension $n+1$. These
are given by the expectation values of $n$ linearly independent
observables that commute with the Hamiltonian, one of these being
the Hamiltonian itself. Writing $E$ for the expectation of ${\hat
H}$, we can then write $\{F_i\}_{i=2 ,\ldots,n}$ for the expectation
values of the other members of the commuting set, which we denote by
$\{{\hat F}_i\}_{i=2 ,\ldots,n}$. By fixing the expectation values
of these conserved quantities we are left with a set of $n$
relative-phase degrees of freedom for the state vector that span the
ergodic subspace of the state space associated with the given
initial state.

We shall show that the equilibrium state corresponds to a uniform
distribution over the toroidal subspace of the quantum state space
spanned by the relative phases. The equilibrium distribution is
characterised, in particular, by a density-of-states function
$\Omega$, which acts as a measure of the size of the toroidal
subspace. The associated density matrix ${\hat\rho}$ is given by the
von Neumann-L\"uders state; that is to say,
\begin{eqnarray}
{\hat\rho}=\sum_{i=0}^n p_i {\hat\Pi}_i, \label{eq:x6}
\end{eqnarray}
where $p_i = |\langle\psi_0| E_i \rangle|^2$. This might be
surprising, since such a state arises most naturally in the context
of measurement theory, where it describes the state of a system
after an energy measurement has been performed. The result is
consistent with the fact that the time average of the dynamics of
the density matrix under unitary evolution is given by the von
Neumann-L\"uders state. It follows that the dynamic average
(\ref{eq:1}) of an arbitrary observable ${\hat O}$ is given by ${\rm
tr}({\hat\rho}{\hat O})$.

To identify the ergodic subspaces of the quantum state space, we
first consider the example of a two-level system, with $n=1$. The
one-parameter family of states generated by unitary evolution can be
written in the form
\begin{eqnarray}
|\psi_t\rangle = \cos\half\,\theta |E_1 \rangle+ \sin\half\,
\theta\,\re^{{\rm i}(\phi+\omega_{10} t)}|E_0\rangle, \label{eq:5}
\end{eqnarray}
where $0\leq\theta\leq\pi$ and $0\leq\phi<2\pi$. The pure state
space has the geometry of a sphere, and unitary evolution gives rise
to a rigid rotation of the sphere around the axis determined by the
two energy eigenstates. Given the initial state $|\psi_0\rangle$,
the dynamical trajectory is the latitudinal circle on which
$|\psi_0\rangle$ lies. The circle is fixed by setting the initial
energy $E$ of the system, which is the only conserved quantity.
Every point on the latitudinal circle is traversed by the dynamical
trajectory, which makes this circle the ergodic subspace of the
state space. The dynamic average of an observable can thus be
replaced by the ensemble average with respect to a uniform
distribution over the circle.

To calculate the associated density of states we compute the
weighted volume in the pure state manifold occupied by the states
having the given property. In general, if we have a set of conserved
quantities $\{G_j\}_{j=1,\ldots,m}$ given by $G_j= \langle\psi_t|
{\hat G}_j |\psi_t \rangle$, then the associated density of states
is
\begin{eqnarray}
\Omega(\{G_j\}) = \int \prod_{j=1}^m \delta(\langle\psi|{\hat
G}_j|\psi \rangle-G_j) \rd V_\psi, \label{eq:6}
\end{eqnarray}
where the integration is over the space of pure states and $\rd
V_\psi$ is the associated volume element. The corresponding
construction for classical systems is considered in \cite{kampen},
where $\Omega(\{G_j\})$ is referred to as a ``substructure
function''. In the case of a two-level system the ergodic circle is
chosen by fixing the expectation of the Hamiltonian: $E=\langle{\hat
H}\rangle$. In terms of the spherical coordinates $(\theta,\phi)$ of
(\ref{eq:5}), the constraint can be written in the form
$(E_1-E_0)\cos^2\half\,\theta=E-E_0$. We thus integrate $\delta(
\cos^2\half\, \theta-(E-E_0)/(E_1-E_0))$ over the pure state
manifold. Since the volume element is $\rd
V=\frac{1}{4}\sin\theta\rd\theta\rd\phi$, the resulting density of
states is
\begin{eqnarray}
\Omega(E) = {\mathds 1}_{\{E_0<E<E_1\}} \frac{\pi}{E_1-E_0}\,,
\label{eq:7}
\end{eqnarray}
where ${\mathds 1}_{\{A\}}$  denotes the indicator function:
${\mathds 1}_{\{A\}}=1$ if $A$ is true and ${\mathds 1}_{\{A\}}= 0$
otherwise.

We proceed to calculate the density of states for $n=2$. In this
case there are two conserved quantities: $E=\langle{\hat H}\rangle$
and $F=\langle{\hat F} \rangle$, where the observable ${\hat F}$
commutes with ${\hat H}$, but is not of the form $a{\hat H}+b{\hat
1}$. The calculation simplifies if we use an equivalent alternative
set of constraints obtained by fixing the expectation values of two
of the energy projectors, say, $p_0=\langle{\hat\Pi}_0 \rangle$ and
$p_1=\langle {\hat\Pi}_1 \rangle$. It follows from the resolution of
identity that $p_2=\langle{\hat\Pi}_2\rangle=1-p_0-p_1$. The unitary
trajectory can be written in the form
\begin{eqnarray}
\fl |\psi_t\rangle = \sin\half
\,\theta_1\cos\half\,\theta_2|E_2\rangle+ \sin\half\, \theta_1\sin
\half\,\theta_2\,\re^{{\rm i}(\phi_1+\omega_{21}t)}|E_1\rangle +
\cos\half\,\theta_1\,\re^{{\rm i}(\phi_2+ \omega_{20}t)}
|E_0\rangle,
\end{eqnarray}
and the two constants of motion are given by
$p_0=\cos^2\half\,\theta_1$ and $p_1=\sin^2\half\,\theta_1\sin^2
\half\,\theta_2$, which fix the variables $\theta_1,\theta_2$.
Therefore, under a generic unitary evolution the ergodic subspace of
the quantum state space is the two-torus ${\mathcal T}^2$ spanned by
$\phi_1,\phi_2$. The density of states is obtained by integrating
$\delta(\cos^2\half\,\theta_1-p_0) \delta(\sin^2\half\, \theta_1
\sin^2\half\,\theta_2-p_1)$ over the pure state manifold, with the
appropriate volume element, which in this case is $\rd V=
\frac{1}{32} \sin \theta_1(1-\cos\theta_1)\sin\theta_2 \rd\theta_1
\rd \theta_2 \rd\phi_1\rd\phi_2$. Performing the relevant
integration we find that $\Omega(p_0,p_1)=\pi^2$ in the triangular
region $\{0<p_0,p_1<1\}\cap\{0<p_0+p_1<1\}$, and vanishes otherwise.

In the case of a general $(n+1)$-level system there are $n$
conserved quantities associated with unitary dynamics. It follows
that under a generic time evolution for which the eigenvalue
differences $\{\omega_{ij}\}$ are incommensurate the typical ergodic
subspace of the quantum state space is given by an $n$-torus
${\mathcal T}^n$. To calculate the density of states $\Omega(p_0,
\cdots,p_{n-1})$ we fix the constraints $\langle{\hat\Pi}_i
\rangle=p_i$ for $i=0,\ldots,n-1$, express these in terms of the
coordinates $(\theta_i,\phi_i)$, and perform the constrained volume
integral over the pure state manifold by using the volume element
\begin{eqnarray}
\rd V = 2^{-n} \prod_{i=1}^n \cos\half\,\theta_i \sin^{2i-1}\half\,
\theta_i \rd\theta_i \rd \phi_i.
\end{eqnarray}
The result is
\begin{eqnarray}
\Omega(p_0, \cdots,p_{n-1}) = \pi^n  \label{eq:11}
\end{eqnarray}
in the hyper-triangular region $\{0<p_0,\ldots,p_{n-1}<1\} \cap
\{0<p_0+\cdots+p_{n-1}<1\}$, and $\Omega(p_0, \cdots, p_{n-1})=0$
otherwise. We see that irrespective of the Hilbert space
dimensionality the density of states is constant in the
hyper-triangular region, and is independent of the energy $E$ and
the conserved quantities $\{F_i\}_{i=2,\ldots,n}$.

The analysis above leads to the following observation. Since for
each $n$ we have identified the ergodic subspaces of the state
space, we are able to apply Birkhoff's theorem to conclude that the
dynamic average of an observable can be replaced by the statistical
average of the observable with respect to an equilibrium state given
by a uniform distribution over the toroidal subspace ${\mathcal
T}^n$.

To compute the expectation of an observable ${\hat O}$ we determine
the density matrix associated with the equilibrium distribution over
the state space. We remark in this connection that the density
matrix associated with a probability density function on the pure
state manifold is the expectation of the pure-state projection
operator with respect to that density function. Now in the energy
basis a pure-state projector can be expressed in the form
\begin{eqnarray}
|\psi\rangle\langle\psi|=\sum_{i,j} \sqrt{p_ip_j}\, \re^{{\rm
i}(\phi_i-\phi_j)}|E_i\rangle\langle E_j|.
\end{eqnarray}
Thus, the diagonal elements $\{p_i\}$ of the pure-state projector
are real, whereas the off-diagonal elements contain phase factors.
The equilibrium distribution has fixed values for the $\{p_i\}$ and
a uniform distribution over the phase variables. It follows that if
we take the average of the projector $|\psi\rangle \langle\psi|$
over the phases, the off-diagonal elements drop out and we are left
with the von Neumann-L\"uders state (\ref{eq:x6}).

The appearance of the von Neumann-L\"uders density matrix as the
equilibrium state is consistent with the fact that the dynamic
average of the density matrix is itself given by the von
Neumann-L\"uders state. This can be seen as follows:
\begin{eqnarray}
\langle{\hat\rho}\rangle &=& \lim_{t\to\infty} \frac{1}{t}
\sum_{i,j} \int_0^t {\hat\Pi}_i {\re}^{-{\rm i}{\hat H}s}
{\hat\rho}_0 {\re}^{{\rm i}{\hat H}s} {\hat\Pi}_j {\rd}s \nonumber
\\ &=& \lim_{t\to\infty}  \frac{1}{t} \sum_{i,j}{\hat\Pi}_i
{\hat\rho}_0 {\hat\Pi}_j \int_0^t {\re}^{-{\rm i}\omega_{ij}s}
{\rd}s \nonumber \\ &=& \sum_i {\hat\Pi}_i {\hat\rho}_0 {\hat\Pi}_i
+ \lim_{t\to\infty} \sum_{i\neq j} {\hat\Pi}_i {\hat\rho}_0
{\hat\Pi}_j \frac{1-\re^{-{\rm i} \omega_{ij}t}}{\ri\omega_{ij}t}
\nonumber \\ &=&  \sum_i {\hat\Pi}_i {\hat\rho}_0 {\hat\Pi}_i =
\sum_i p_i {\hat\Pi}_i. \label{eq:13}
\end{eqnarray}
In particular, we see that the timescale involved for the
averaging to become effective is determined by the energy
differences. We thus conclude that the dynamic average of an
observable ${\hat O}$ is given by ${\rm tr}({\hat\rho}{\hat O})$,
where ${\hat\rho}$ is given by (\ref{eq:x6}). This representation
of the density matrix, however, does not make the thermodynamic
properties of the equilibrium state immediately apparent. We shall
demonstrate, however, that in association with the conserved
quantities $(E,\{F_i\})$ there is a corresponding system of
conjugate variables $(\beta,\{\mu_i\})$ that can be given a
consistent thermodynamic interpretation. In the case of the energy
the conjugate variable has the interpretation of the inverse
temperature. For the other observables the associated conjugate
variables can be interpreted as chemical potentials. This suggests
that the equilibrium state arising from unitarity and ergodicity
might be of a grand canonical type. The conjugate variables are
defined as follows. Writing (\ref{eq:x6}) for the density matrix
associated with the toroidal subspace characterised by the
conserved quantities $(E,\{F_i\})$ we have ${\rm tr}{\hat\rho}=1$,
${\rm tr}({\hat\rho}{\hat H})=E$, and ${\rm tr}({\hat\rho}{\hat
F}_k)=F_k$. Let us define a family of $n+1$ operators $\{{\hat
G}\}_{i=0,1,\ldots,n}$ by setting ${\hat G}_0= {\hat 1}$, ${\hat
G}_1= {\hat H}$, and $\{{\hat G}_i\}_{i=2,\ldots ,n} = \{{\hat
F}_i\}_{i=2,\ldots ,n}$, writing $\{ G\}_{i= 0,1,\ldots,n}$ for
the corresponding expectation values with respect to ${\hat\rho}$,
so $G_0=1$, $G_1=E$, and $\{G_i\}_{i=2,\ldots ,n} =
\{F_i\}_{i=2,\ldots ,n}$. In other words, ${\rm
tr}({\hat\rho}{\hat G}_i) = G_i$ for $i=0,1,\ldots,n$. It follows
from (\ref{eq:x6}) that
\begin{eqnarray}
\sum_{i=0}^n p_i\, {\rm tr}\left({\hat\Pi}_i{\hat G}_j\right) = G_j.
\end{eqnarray}
Thus, writing $g_{ij}={\rm tr}({\hat\Pi}_i{\hat G}_j)$ and defining
$h_{jk}$ by $\sum_{j=0}^n g_{ij}h_{jk}=\delta_{ik}$, we see that
\begin{eqnarray}
p_k = \sum_{j=0}^n G_j h_{jk},  \label{eq:zz0}
\end{eqnarray}
and therefore that
\begin{eqnarray}
{\hat\rho} = \sum_{j,k=0}^n G_j h_{jk} {\hat\Pi}_k. \label{eq:zz1}
\end{eqnarray}
To verify that $h_{jk}$ exists, we observe that if it did not, then
there would exist a nonzero vector $\xi_i$ such that $\sum_{j=0}^n
g_{ij} \xi_j=0$; but that would imply ${\rm tr}
({\hat\Pi}_i\sum_{j=0}^n {\hat G}_j\xi_j)=0$ for all $i$, and hence
$\sum_{j=0}^n {\hat G}_j \xi_j=0$, contrary to the assumption that
the ${\hat G}_j$ are linearly independent.

Formula (\ref{eq:zz1}) gives ${\hat\rho}$ as a function of $E$ and
$\{F_i\}$. Therefore, writing $S=-{\rm tr}({\hat\rho}\ln{\hat\rho})$
for the entropy, we obtain an expression for $S$ as a function of
$E$ and $\{F_i\}$. The associated conjugate variables are then
defined by the thermodynamic relation
\begin{eqnarray}
\rd S=\beta \rd E + \sum_{k=2}^{n} \mu_k \rd F_k, \label{eq:14}
\end{eqnarray}
where $\beta$ is the effective inverse temperature and $\{\mu_k\}$
are the effective chemical potentials. This shows, on account of the
linear independence of the observables, the equivalence of the
specification of either (i) the initial state $|\psi_0\rangle$ up to
relative phases, (ii) the probabilities $p_i=|\langle\psi_0
|E_i\rangle|^2$, (iii) the expectation values $E$ and $\{F_i\}$, or
(iv) the conjugate variables $\beta$ and $\{\mu_i\}$. We can
therefore investigate how the equilibrium density matrix
(\ref{eq:x6}) can be expressed either in terms of the extensive
variables $E$ and $\{F_i\}$, or in terms of the conjugate variables
$\beta$ and $\{\mu_i\}$.

For the various representations of the density matrix we consider
first the example of the two-level system. In this case we solve the
relations $p_0+p_1=1$ and $p_0E_0+p_1 E_1=E$ for the diagonal
elements $p_0,p_1$ of ${\hat\rho}$, and obtain
\begin{eqnarray}
{\hat\rho}(E) = \left( \begin{array}{cc} \frac{E_1-E}{E_1-E_0} & 0
\\ 0 & \frac{E-E_0}{E_1-E_0} \end{array} \right). \label{eq:19}
\end{eqnarray}
Computing the entropy and using the relation $\rd S=\beta\rd E$ we
can express the inverse temperature as a function of $E$. The result
is
\begin{eqnarray}
\beta(E) = \frac{1}{E_1-E_0} \ln \left( \frac{E_1-E}{E-E_0} \right)
. \label{eq:20.0}
\end{eqnarray}
By inverting this relation, we then obtain
\begin{eqnarray}
E(\beta) = \frac{E_0\re^{-\beta E_0}+E_1\re^{-\beta E_1}}
{\re^{-\beta E_0}+\re^{-\beta E_1}}. \label{eq:20}
\end{eqnarray}
Expression (\ref{eq:20}) is, however, the expectation of the energy
with respect to the canonical density matrix. That is to say,
(\ref{eq:19}) can be expressed in the form
\begin{eqnarray}
{\hat\rho}(E) = \frac{1}{Z(\beta)}\left( \begin{array}{cc}
\re^{-\beta E_0} & 0 \\ 0 & \re^{-\beta E_1} \end{array} \right),
\label{eq:21}
\end{eqnarray}
where $Z(\beta)=\re^{-\beta E_0}+\re^{-\beta E_1}$. The important
point here is that the inverse temperature $\beta$ is not specified
exogenously via the introduction of a heat bath. Rather, it is
defined endogenously, through the specification of the energy of the
equilibrium state associated with the given initial state.

Let us now turn to the proof of the quantum ergodic theorem in the
general case. It follows from (\ref{eq:zz1}) that the entropy is
given by
\begin{eqnarray}
S = -\sum_{k=0}^n \left(\sum_{j=0}^n G_j h_{jk}\right) \ln
\left(\sum_{j=0}^n G_j h_{jk}\right).
\end{eqnarray}
Thus, defining $\gamma_i=\partial S/\partial G_i$ by use of this
expression, we find that
\begin{eqnarray}
\gamma_i = - \sum_{k=0}^n h_{ik} \left[ \ln\left(\sum_{j=0}^n G_j
h_{jk}\right)+1\right] = - \sum_{k=0}^n h_{ik} \left( \ln
p_k+1\right), \label{eq:x21}
\end{eqnarray}
by (\ref{eq:zz0}), and hence
\begin{eqnarray}
\ln p_i + 1 &=& -\sum_{j=0}^n g_{ij} \gamma_j = -\sum_{j=0}^n {\rm
tr} \left({\hat\Pi}_i{\hat G}_j\right) \gamma_j \nonumber \\ &=&
-\gamma_0 - \gamma_1 {\rm tr} \left( {\hat\Pi}_i{\hat H} \right) -
\sum_{j=2}^n \gamma_j {\rm tr} \left( {\hat\Pi}_i{\hat F}_j \right).
\label{eq:x22}
\end{eqnarray}
Setting $\gamma_1=\beta$ and $\{\gamma_i\}_{i=2,\ldots,n}=\{ \mu_i
\}_{i=2,\ldots,n}$, these relations are then sufficient to
determine the diagonal elements $\{p_i\}_{i=0,\ldots,n}$ of the
equilibrium density matrix in terms of the intensive variables,
and we are led to the grand canonical ensemble (\ref{eq:2}) with
the identification $\gamma_0=\ln Z-1$. The effective inverse
temperature, however, is not associated with an external heat
bath, but rather is intrinsic to the system, and a similar remark
applies to the effective chemical potentials. The fact that the
conjugate variables are determined endogenously shows that our
result does not require an assumption of entropy maximisation.

In the case of a degenerate Hamiltonian, the ergodic subspace of the
state space is contracted to a smaller torus ${\mathcal
T}^m\subset{\mathcal T}^n$, where $m+1$ is the number of distinct
energy eigenvalues. This follows from the fact that since some of
the eigenvalue differences $\omega_{ij}$ vanish, only $m$ of the $n$
relative phases for the unitary trajectory $|\psi_t\rangle$ vary in
time. As a consequence, we need only to consider $m-1$ independent
observables $\{{\hat F}_i\}$ whose eigenspaces coincide with that of
the Hamiltonian. In other words, there are only $m$ terms, given by
${\hat H}$ and $\{{\hat F}_i\}_{i=2,\ldots,m}$, in the exponent of
(\ref{eq:2}) for the grand canonical density matrix. As an example
consider the case of a three-dimensional Hilbert space where the
energy eigenvalues are given by $E_0$, $E_1$, and $E_1$. The
elements of the density matrix are $p_0=(E_1-E)/(E_1-E_0)$ and
$p_1=p_2= (E-E_0)/2(E_1-E_0)$. A short calculation making use of the
relation $\rd S= \beta\rd E$ then shows that
\begin{eqnarray}
E(\beta) = \frac{E_0\re^{-\beta E_0}+2E_1\re^{-\beta E_1}}
{\re^{-\beta E_0}+2\re^{-\beta E_1}}, \label{eq:23}
\end{eqnarray}
which is evidently the expectation of ${\hat H}$ with respect to the
canonical density matrix ${\hat\rho}=\exp(-\beta{\hat H})/{\rm tr}
\exp(-\beta{\hat H})$.

A challenging open issue is to understand the implications of the
quantum ergodic theorem for macroscopic systems. In the case of a
large quantum system the energy spectrum of a typical model
Hamiltonian is highly degenerate. As a consequence, the number of
independent macro-observables ${\hat H}$ and $\{{\hat F}_i\}$
required for the exact specification of the equilibrium density
matrix is significantly reduced. For real systems, however, due to
the complexity of internal interactions one would expect the
degeneracies in model Hamiltonians to split into closely located
but distinct levels. Therefore, the specification of a small
number of macro-variables will only provide an approximate
description of the equilibrium state for real systems. On the
other hand, if there are large clusters of observables with the
property that in the equilibrium state defined by (\ref{eq:2}) the
chemical potentials are approximately equal, then the resulting
state can be adequately characterised by a small number of
macro-variables, and thus can be regarded as effectively
classical. It is interesting in this connection to contrast the
results obtained here for quantum systems with the corresponding
results for strictly classical systems: While ergodicity is
generic for quantum systems, classically it is
exceptional~\cite{markus}. The fact that the characterisation of
the equilibrium state of a quantum system is simpler, and that the
equilibrium distribution can be derived dynamically by use of an
ergodicity argument, might be related to the special structures of
energy surfaces in quantum phase spaces~\cite{brody}.

\vskip4pt DCB acknowledges support from The Royal Society.

\end{document}